\documentclass[superscriptaddress,prl,twocolumn,amsmath,amssymb,floatfix,showpacs]{revtex4-1}

\usepackage{graphicx}
\usepackage{bm}
\usepackage{amssymb}
\usepackage{color}
\usepackage{epsfig}
\usepackage{subfigure}

\usepackage{amsmath}
\usepackage{hyperref}
\usepackage{bbold}
\hypersetup{
    pdfstartview={FitH},
   pdfpagemode={None}
   }
\newcommand{\be}{\begin{equation}}
\newcommand{\ee}{\end{equation}}
\begin{document}

\title {Cell size regulation in microorganisms}

\author{Ariel Amir}
\affiliation {Department of Physics, Harvard University, Cambridge, MA 02138, USA}

\begin {abstract}
 Various rod-shaped bacteria such as the canonical gram negative \emph{Escherichia coli} or the well-studied gram positive \emph{Bacillus subtilis} divide symmetrically after they approximately double their volume. Their size at division is not constant, but is typically distributed over a narrow range. Here, we propose an analytically tractable model for cell size control, and calculate the cell size and inter-division time distributions. We suggest ways of extracting the model parameters from experimental data. Existing data for \emph{E. coli} 
 supports \emph{partial} size control, and a particular explanation: a cell attempts to add a \emph{constant volume} from the time of initiation of DNA replication to the next initiation event. This hypothesis explains how bacteria control their tight size distributions and accounts for the experimentally observed correlations between parents
and daughters as well as the exponential dependence of size on growth rate.
\end {abstract}


 \maketitle
Microorganisms such as bacteria 
come in a diverse set of shapes and sizes. Nonetheless, individual strains have remarkably reproducible shapes, and a narrow distribution of sizes \cite{cooperb,koch,jun, dekker2}. Many bacteria, such as \emph{E. coli}, are rod-shaped, and during their exponential growth phase they elongate while maintaining a constant diameter. After approximately doubling their length (as well as mass and volume), and completing DNA replication for their offspring, they divide symmetrically into two approximately identical daughter cells.  In spite of decades of research, we still do not have a good understanding of how cells regulate their shape, both mechanically (i.e., what is the biophysical feedback necessary to achieve a rod-shape cell? \cite{perspectives}) and dimensionally: the coefficient of variation (standard deviation:mean, CV) can be as low as 0.1 for bacteria \cite{koch}. Bacteria are also remarkable in their ability to have a generation time that is shorter than the time it takes them to replicate DNA: doubling time $\tau_d$ for \emph{E. coli} in rich media at $37^\circ$C is about $20$ mins, while $T_r \approx 60$ mins are needed from initiation of DNA replication to cell division. This apparent paradox is explained by the existence of \emph{multiple replication forks}: in these situations, a cell will already start replicating DNA for its 4 granddaughters (or 8 great-granddaughters), in order for the replication to complete in time.

Many models for cell size regulation exist in the literature \cite{koch, fantes, levin, review1,jorgensen1,cellcyclebook,cooperb,size1, size2}. Different strategies will yield  particular cell size and inter-division time distributions, as well as distinct correlations. Hence, it is important to understand the connection between different regulation models and the resulting distributions and correlations. Moreover, there are two seemingly contradictory results in the literature: the first is the model by Donachie \cite{donachie}, which shows that the measured exponential dependence of bacterial size on growth rate \cite{schaechter} is consistent with initiation of DNA replication at a constant, growth-rate-independent volume per replication fork -- suggesting a mechanistic picture in which a cell ``knows" of its size and initiates replication when reaching a critical one. This model would imply that size at birth and division would not be correlated: since the time from initiation to division is constant \cite{helmstetter}, the size at division will be independent of the size at birth. However, experiments show that there are strong correlations between the two \cite{correlation}.

We will show here how these two results can be elegantly reconciled within a minimal model, which will be analytically tractable. We will suggest a mathematical framework which is able to capture and extend several existing models, and will use it to analyze the correlations and cell size distributions. We shall show that the aforementioned experimental data for \emph{E. coli} supports a mechanism of cell size regulation in which the cell attempts to \emph{add a constant volume from the event of initiation of DNA replication to the next initiation event} \cite{incrementalmodel}. This model will be consistent with the results discussed in Ref. \cite{donachie}, predicting an exponential dependence of cell size on growth rate, but will also quantitatively account for the positive correlations between size at birth and division \cite{correlation} and negative correlations between size at birth and inter-division time \cite{size2}. We will show that for size-additive noise the size distribution is Gaussian, while for time-additive (i.e., size-multiplicative) noise the resulting size distribution is log-normal -- and hence right-skewed. As shown in the Supplementary Information (SI), experimentally measured distributions are indeed skewed, and for this reason we focus on the analysis of time-additive noise in the main text and defer the size-additive case to the SI. The standard deviations of both size and inter-division time distributions are controlled by a \emph{single} parameter.



The tools which we shall use will parallel those used when solving problems in statistical mechanics. Multiplicative noise and the log-normal distributions which emerge from our model also occur frequently in a broad class of problems in physics, such as relaxations in glasses \cite{amir_glass_PNAS}, intensity fluctuations in lasers \cite{laser} and the modelling of financial markets \cite{econophysics_book, bouchaud_potters}. However, in contrast to most physical systems, negative feedback (i.e., control) is a necessary feature of biological systems, including the problem studied here. In many of these examples, a Langevin approach is used to model the stochastic process \cite{langevin}, with either additive or multiplicative noise. As is also exemplified in our work, multiplicative noise is known to lead to distinctively different distributions than additive noise, in certain cases leading to non-equilibrium phase transitions \cite{multiplicative1} and power-law tails \cite{multiplicative2}.

\emph{Exponential growth of a single cell and regulation models.-} The question of the mode of growth of a single bacterium has been a long standing problem, with linear and exponential growth the most common models considered \cite{cooperb,koch, cooper}. Recent experiments show that individual cell volume grows exponentially, for various bacterial strains \cite{jun,jacobs,godin,golding}. 
In fact, if cells grow at a rate that is proportional to the amount of protein they contain \cite{hwa, amir_nelson_pnas}, as long as the protein concentration is constant, the cells will grow exponentially in mass and volume. We shall assume exponential growth of \emph{volume} throughout this paper, $v(t) \propto 2^{t/\tau_d}$, and neglect fluctuations in the growth rate. Furthermore, when discussing bacterial division we will assume that cells divide precisely in half since experimental results \cite{marr2} show that division occurs at the mid-cell to an excellent approximation (this assumption is not justified for budding yeast, which divides asymmetrically). 

Cells need a feedback mechanism that will control their size distribution. If cells grew for a constant time $t=\tau_d$, random fluctuations in the timing would make the size of the cells at division, $v_{d}$, perform a random walk on the volume axis, and thus this mechanism does not control size. Another regulatory strategy is that of division at a critical mass, or of initiation of DNA replication at a critical size. These ideas are prevalent in the literature \cite{koch,cooperb}, but we will show that existing experimental data for \emph{E. coli} argue against them. We shall consider the following class of models: upon being born at a size $v_{b}$, the cell would ideally attempt to grow for a time $\tau(v_b)$ such that its final volume at division is $v_d= f(v_{b})$. If the function $f(v_b)=const$, we are back to the critical mass model. The constant time model can also be cast in this language: since the growth is exponential, attempting to grow for a prescribed, constant time $\tau_d$ is the same as having $f(v_b)=2 v_b$. Another important model that has been suggested is the so-called ``incremental model", in which the cell attempts to add a constant volume $v_0$ to its newborn size \cite{comment_incremental}. In this case: $f(v_b)= v_b + \Delta $. In the following, we suggest a method through which an arbitrary regulatory model described by a function $f(v_b)$ can be approximately solved, i.e., we can find all the involved distributions analytically, finding excellent agreement with the numerically exact solutions. We also provide methods to extract the model parameters from experimental data. 

Our calculations can be done either for time-additive or size-additive noise. We will show that time-additive noise leads to approximately log-normal size distributions, while size-additive noise gives Gaussian distributions. Hence, measuring the distribution skewness is a useful way to distinguish between the two cases. In the SI we show that experiments on \emph{E. coli} agree better with a time-additive noise, and for this reason we focus on this case here, and defer the calculations of the case of size-additive noise to the SI.
%

\emph{The model.-} We assume that the cell attempts to divide at a volume $v_d= f(v_{b})$, as previously explained, by attempting to grow for the appropriate amount of time $t_{a}$ which is a function of $v_d$. We assume that to this time is added a random noise $t_{n}$, which we assume to be Gaussian. The magnitude of this noise will dictate the width of the resulting size and inter-division time distributions. Thus we have:

\be t_{\rm{growth}} = t_{a}+t_{n}= \tau_d \log_2[f(v_b)/v_b]+t_{n}, \label{model0} \ee with $t_{n}$ assumed to be a random variable with: $P(t_{n}) =\frac{1}{\sqrt{2\pi \sigma_T^2}} e^{-\frac{t_{n}^2}{2 \sigma_T^2}}.$ The model is similar to that discussed in Ref. [\onlinecite{siggia}], where the molecular mechanisms leading to the noise in budding yeast are studied.

We will calculate the inter-division time and volume distributions. The key insight is that for noise that is not too large (equivalent to size distributions which are not too broad, i.e., with a small CV), it is the behavior of $f(v_b)$ around the average newborn size $v_0$ that is the most important. Therefore we can Taylor expand $f(v_b)$ around $v_0$:

\be f(v_b) \approx f(v_0)+f'(v_0)(v_b-v_0). \label{regulate} \ee As an example, the incremental model has $f'(v_0)=1$ and $v_0=\Delta$, while the critical size model has $f'(v_0)=0$.

Any two models that agree to lowest order, will result in similar distributions -- provided the noise is not too large. We therefore choose to solve an equivalent model, that will be amenable to analytic treatment, and that can be viewed as an interpolation between the critical size model and the constant doubling time model. We choose:

\be t_{a} = \tau_d [1 + \alpha \log_2(v_0/v_b)], \label{model1}\ee
which corresponds to the regulatory function: $f(v_b) = v_0 2^{t_{a}} = 2 v_b^{1-\alpha} v_0^\alpha$. The case $\alpha=0$ corresponds to constant doubling time model ($f'(v_0)=2$), while $\alpha=1$ corresponds to the critical size model ($f'(v_0)=0$). Importantly, for $\alpha=1/2$ we have $f'(v_0)=1$, as does the incremental model: hence, using a target function like this gives results close to a perfect realization of the incremental mode.

\emph{Solution of size and  inter-division time distributions.-} We shall 
consider the case of \emph{symmetric} division, relevant for many rod-shaped bacteria. For a newborn size $v_b$, we have for the next newborn volume: $v^{new}_b= v_0^\alpha v_b^{1-\alpha} 2^{t_{n}/\tau_d}$.

Therefore:
\be \log_2(v^{new}_b/v_0)= (1-\alpha)\log_2(v_b/v_0) + t_{n}/\tau_d, \label{corr} \ee

From stationarity of the stochastic process we know that $P(v^{new}_b) = P(v_b)$. Since $t_{n}$ is a Gaussian variable, we find that $\log_2(v_b)$ is also a Gaussian variable, and hence $P(v_b)$ would be a \emph{log-normal} distribution. If we denote the variance of $\log_2(v_b/v_0)$ by $\sigma^2_v$, we have $\sigma^2_v = \sigma^2_v(1-\alpha)^2 + \frac{{\sigma_T}^2}{{\tau_d}^2}$, therefore the newborn size distribution is:

\be P(v_b) =\frac{1}{\sqrt{2 \pi} \ln(2)  \sigma_v}  \frac{e^{-\frac{[\log_2(v_b/v_0)]^2}{2 \sigma^2_v}}}{v_b}, \label{newsize} \ee
with
\be  \sigma^2_v= \frac{{\sigma_T}^2}{\tau_d^2 \alpha(2-\alpha)}. \label{std_l} \ee

Note that the average cell size is $\bar{v}=v_0 e^{\sigma^2_v/2}$; for realistic values of $\sigma_v$ it will only be a few percent larger than $v_0$. Similarly, the standard deviation of the size distribution will be approximately $\sigma_{s}\approx \log(2) \sigma_v v_0$, and the coefficient of variation is thus: $v_{CV} \approx \log(2) \sigma_v $. The skewness of the distribution is positive: $\gamma_1 \approx 3 \log(2) \sigma_v$, and provides a useful test of the assumption of a time-additive rather than size-additive noise, as we elaborate on in the SI.

We can now find the distribution of division times using: $t_{d}=t_{a}+ t_{n}$. Since $v_b$ depends only on the noise of previous generations, $t_{a}$ is independent of $t_{n}$, and since $\log_2(v_b/v_0)$ and $t_{n}$ are Gaussian variables, the resulting inter-division time distribution is also Gaussian, and has a variance given by:

\be Var[t_d]= \tau_d^2\alpha^2 \sigma^2_v + {\sigma_T}^2 ={\sigma_T}^2 \frac{2}{2-\alpha}.  \label{time_std} \ee

 In the case $\alpha \rightarrow 0$, we find that $\sigma_v$ diverges (an extremely broad distribution of newborn sizes), but the  inter-division time distribution is narrow: $Var[t_d]\rightarrow {\sigma_T}^2$, as should clearly be the case since there is no size feedback mechanism in this case. Note that for any positive value of $\alpha > 0$ there will be a stationary size distribution, but for $\alpha=0$ there is no stationary distribution.

From Eq. (\ref{time_std}) we find that the CV of the distribution of inter-division times is given by: $t_{CV}= \frac{\sigma_T}{\tau_d} \sqrt{\frac{2}{2-\alpha}}.$ It is instructive to consider the dimensionless quantity:

\be \gamma \equiv v_{CV}/ t_{CV} \approx \frac{\log(2)}{\sqrt{2\alpha}}.\label{gamma}\ee By constructing $\gamma$ from the experimental distributions we can extract the value of $\alpha$ and find the form of size regulation utilized by the organism, if the division is symmetric. Later we shall show an additional, independent way of extracting $\alpha$, which will be more robust against measurement noise since it will rely on correlations rather than the distribution widths. 

%

\begin{figure}
\centering
\mbox{\subfigure{\includegraphics[height=0.4\linewidth]{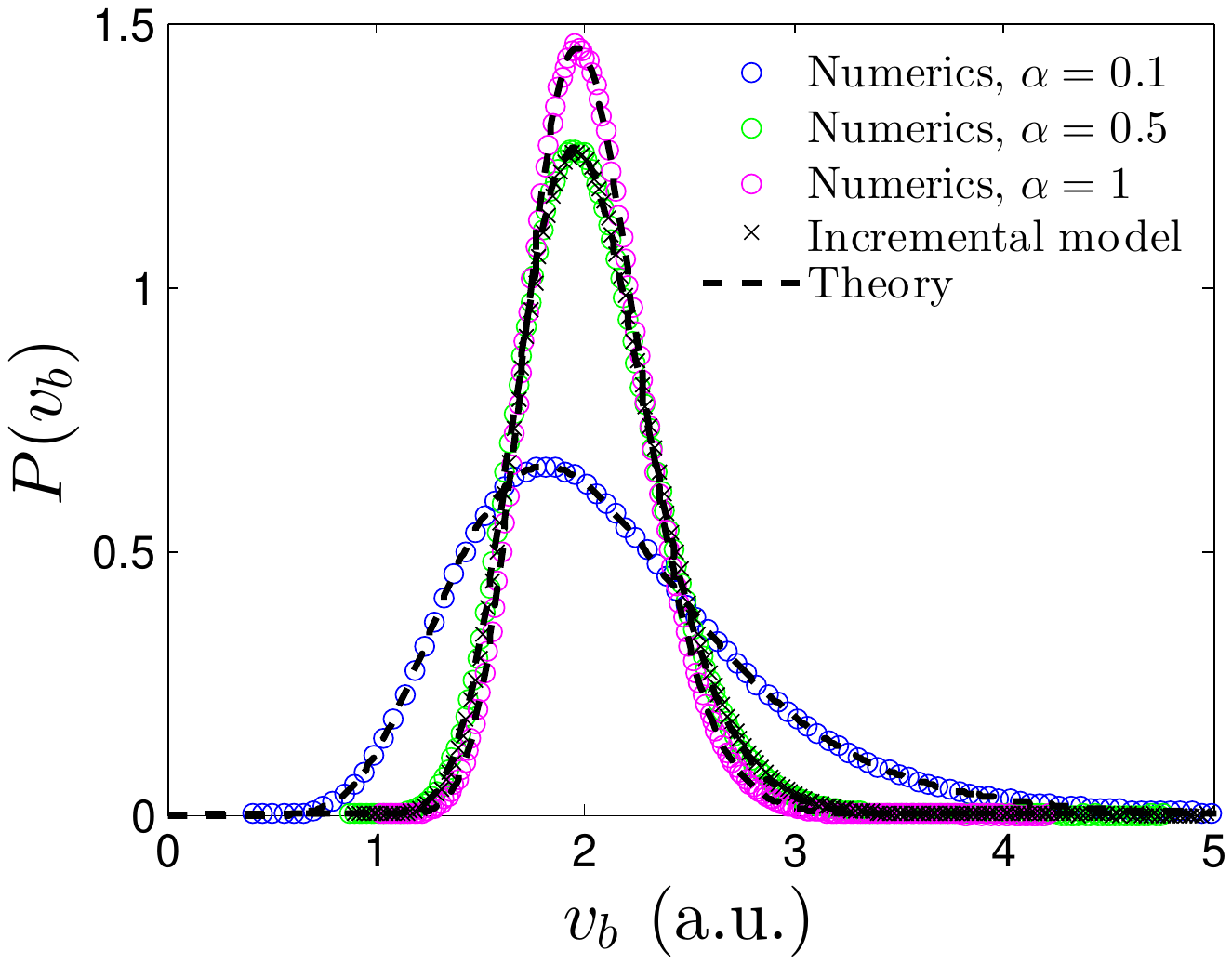}}\quad
\hspace{-.5 cm} \subfigure{\includegraphics[height=0.4\linewidth]{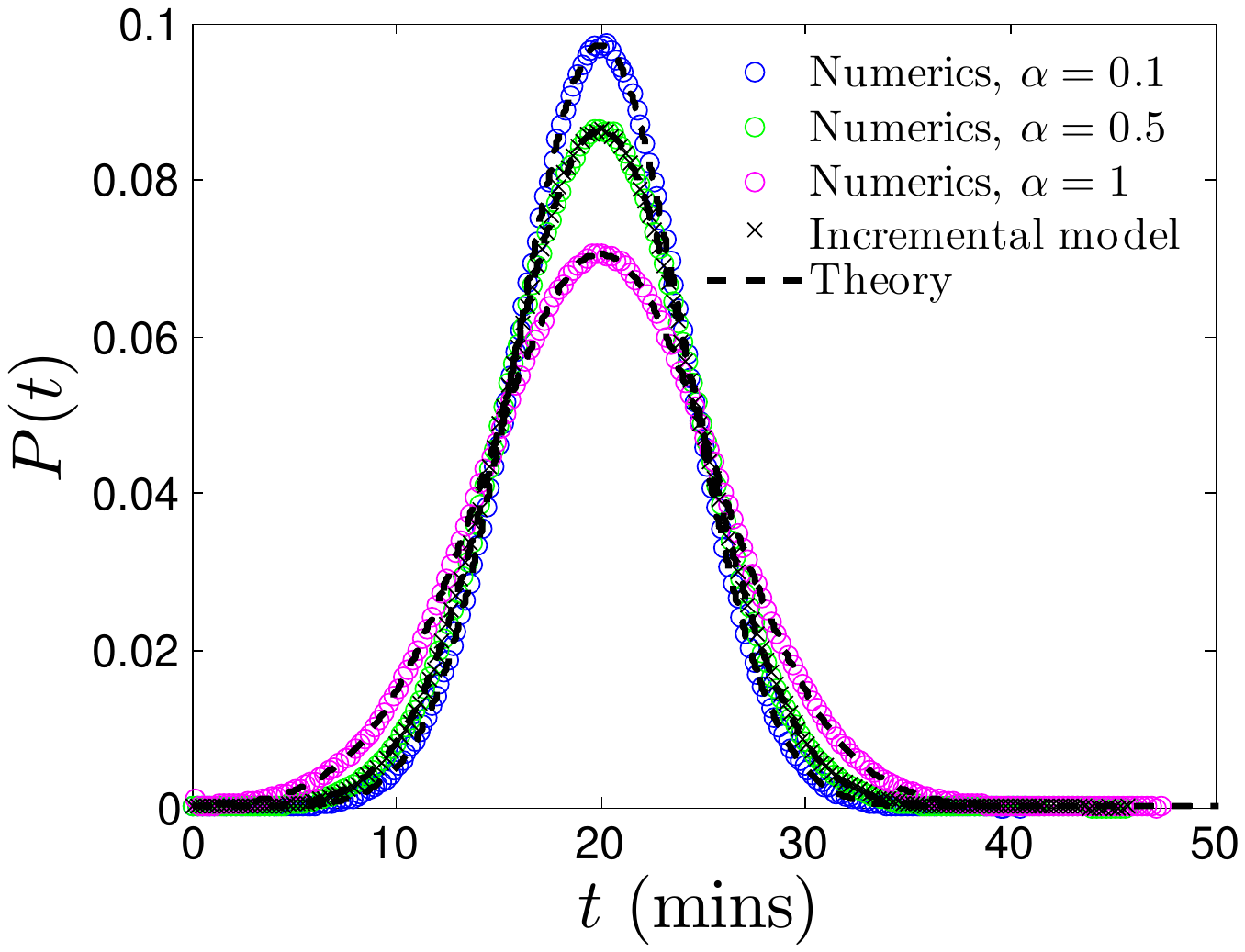} }}
\caption{Comparison between the analytical results of the model for varying values of $\alpha$ (Eqs. (\ref{newsize}-\ref{time_std})), and numerics. Choosing $\alpha=1/2$ provides an excellent approximation for the incremental model, as the effective size regulation of the two models agrees to lowest order. The parameters of the model are chosen according to their realistic values for \emph{E. coli} growing at $37^o$: doubling time is $\tau_d=20$ mins and $\sigma_T/\tau_d=0.2$ \cite{koch}. For each case, the numerical distribution is extracted from a sequence of $10^7$ divisions. }
\label{numerics}
\end{figure}

Fig. \ref{numerics} compares the numerically obtained size distribution for various values of $\alpha$ and the incremental model, with the result of Eq. (\ref{newsize}), finding excellent agreement. In the SI we extend this comparison to various noise magnitudes. Our model captures the numerically exact solution very well and Eq. (\ref{model1}) provides a useful tool to capture a generic division strategy characterized by an arbitrary function $f(v_b)$.

\emph{Extracting the parameters from experiments.-} Within the class of models proposed here, the value of $\alpha$ can be obtained by considering the correlations between size at birth and size at division. For a narrow size distribution, performing a linear regression analysis between the size at birth and the size at division would not be very different than doing the regression between $x=\log_2(v_b/v_0)$ and $y=\log_2(v_d/v_0)$ (for the data corresponding to Fig. 1, for example, the difference between the two methods is less than one percent). Within our model, the value of the dimensionless slope $\beta$ can be readily found from Eq. (\ref{regulate}): since the noise is uncorrelated with the random variable $x$, we have $\beta = f'(v_0)=2(1-\alpha)$ (i.e., it is 1 for the incremental model). For symmetric divisions, the slope of the linear regression between a newborn cell and the size of the daughter cell immediately after division would thus equal $1-\alpha$. The coefficient of correlation between the newborn cell and the newborn daughter cell is: $ C = \mathbb{E}(x y)/\sigma_x \sigma_y = (1-\alpha)\mathbb{E}(x^2)/\sigma_x \sigma_y. \label{cov}$ From stationarity, we know that the distribution of $v_b$ and of $v^{new}_b$ are identical, hence $\sigma_y=\sigma_x$, and: $ C = 1-\alpha.$ Therefore, for the incremental model the correlation coefficient between mother and daughter cells should be 0.5.


Upon fixing the value of $\alpha$, 
a \emph{single} parameter, $\sigma_T$, will determine the distributions of both size and division time, and the calculations performed here would allow one to scale both distributions using this single parameter. For the time-additive noise analyzed here, the model predicts an approximately log-normal newborn size distribution, given by Eq. (\ref{newsize}), and a Gaussian inter-division time distribution, given by Eq. (\ref{time_std}), whose standard deviation is larger than $\sigma_T$. In the SI we show that for size-additive noise, one obtains a skewed time-distribution but a Gaussian size distribution -- in contrast to what is observed experimentally \cite{size2}. Therefore, observing the distribution shape provides useful information regarding the source of the noise. Further experiments are needed to elucidate the molecular source of this multiplicative noise.


\emph{Cell size control in E. coli.-}
Experimentally, various correlation coefficients were measured for \emph{E. coli} at slow growth conditions in Ref. [\onlinecite{correlation}], using the membrane elution technique. The correlation coefficient between newborn cell and daughter cell size was found to be $C=0.55$, close to the theoretical $1/2$ value expected for the incremental model. There was a strong correlation (0.8) between size at initiation of DNA replication and size at division, as we would expect from the assumption of exponential growth and that the time from initiation to division is constant \cite{helmstetter}. Furthermore, the CV of their measured size distribution is smaller than that of the inter-division time distribution by $\gamma \approx \log(2)$, as we expect for the incremental model from Eq. (\ref{gamma}). 
 Yet these observations appear to be in direct contradiction to the idea that initiation occurs at a critical size \cite{donachie}. The key point is that the experiments only show that there is a critical size for initiation of DNA replication (independent of growth rate), \emph{on average}. It is only from the fluctuations (i.e., correlations) that one can understand whether the underlying regulatory mechanism utilizes a critical size or integrates volume -- as we shall propose is the case. Ref. \cite{incrementalmodel} gives a simple biophysical implementation of the incremental model, which will reconcile these seemingly contradictory results and will realize a particular case of the class of model we proposed here: in this model, a protein $A$ is forced to have a growth-rate-independent density throughout the cell using a negative feedback in its regulation, and a second protein $B$ is produced whenever $A$ is. In this way when cell volume grows (and only then), more $A$ and $B$ proteins are generated in an amount proportional to the change in \emph{volume}. The hypothesis is that $B$ proteins localize at their potential initiation site (which we will assume to be one of the replication origins), and only when their total \emph{number} at each origin reaches a critical value does initiation of DNA replication occur, after which $B$ is degraded. Note that two types of proteins are necessary, since in order to measure volume differences $A$ must be spread throughout the cell, while $B$ has to localize to measure an absolute number (rather than concentration). See Ref. \cite{incrementalmodel} for further details.   

 We shall now show that this model realizes the incremental model (corresponding to Eq. (\ref{model1}) with $\alpha=1/2$), yet with a $\Delta$ which depends on size in a particular way: Let us assume that that there are $2^n$ replication forks at work (and hence $2^{n+1}$ replication origins), and that initiation of DNA replication occurred at a volume $v_i$ for one of them. Protein $B$ will be accumulated at each origin until a critical amount is reached. This implies that the next initiation will occur (on average) at a volume $v^{next}_i=(v_i + 2^{n+1} \Delta)/2$, where according to the above biophysical mechanism $\Delta$ is \emph{independent} of the growth rate. According to our assumption, there will be $n$ division events from initiation to the end of replication. Since growth is exponential and we are assuming perfectly symmetric divisions, if the cell volume at initiation is $v_i$ its volume at the end of DNA replication is $v^d=v_i 2^{T_r/\tau_d-n}$, regardless of when the intermediate division events happened. Its daughter cell will be born with a volume $v_{nb}=v^d/2$, and its size at division will equal, by the same reasoning:
 \be V^{next}_d=v^{next}_i 2^{T_r/\tau_d-n} = \frac{v_i}{2} 2^{T_r/\tau_d-n} + 2^{n} \Delta 2^{T_r/\tau_d-n}.\ee
 Thus we have (up to the noise):
 \be v^{next}_d= v_{nb} + \hat{\Delta}, \label{ecoli} \ee
with $\hat{\Delta} \equiv \Delta 2^{T_r/\tau_d}$. This corresponds to Eq. (\ref{regulate}) with $v_0=\hat{\Delta}$ and $f'(v_0)=1$, and according to our results the average cell size will be $\hat{\Delta}$ -- in agreement with the experimental results seeing precisely this exponential dependence of bacterial size on growth rate, with $T_r$ the exponent \cite{schaechter}. This model naturally accounts for the ``quantization" of the cell critical mass at initiation at different growth rates \cite{donachie}, without necessitating the measurement of an absolute mass or volume. Moreover, it is plausible that the source of noise in adding the incremental volume will be due to ``molecular noise" (number fluctuations of protein $B$), and would therefore be weakly dependent on growth rate. The same reasoning which leads to Eq. (\ref{ecoli}) would suggest that $\sigma_T$ (the noise standard deviation) should depend on the growth rate in the same exponential way as $\hat{\Delta}$. This implies that the CV of size distributions should be weakly dependent on growth rate (see Eq. (\ref{std_l})), an expectation supported by Ref. \cite{trueba}.

Thus, we have shown that using our calculations and the interpretation in terms of the incremental model explains various experimental results. In fact, the model also makes precise predictions with regards to additional correlations: for example, it is possible to show that for the incremental model the size correlation coefficient between cells $N$ generations apart is $2^{-N}$. Similarly, the model predicts a negative correlation between the size at birth and the inter-division time, as expressed by Eq. (\ref{model1}) for $\alpha=1/2$. Very recently, novel analysis of experimental data was published, which studied precisely this correlation \cite{size2}. Doumic \emph{et al.} find, based on two different experimental systems, a coefficient of correlation of -0.5: exactly as predicted by our model. This gives a particularly simple and transparent interpretation to their analysis, and provide additional, strong support for the incremental model. Refs. [\onlinecite{elf, size1}] find similar negative correlations between newborn size and inter-division size, supporting our conclusion.



All of these provide additional support for the relevance of this model to cell size control in \emph{E. coli}, and most likely to other organisms as well. It is possible, however, that alternative biophysical mechanisms may lead to the same correlations and size dependencies calculated here, and for this reason finding the underlying biological mechanism is important; In recent years, dnaA has been shown to have properties reminiscent of the biophysical model described here \cite{levin}, where its active and inactive forms correspond to the roles of proteins $A$ and $B$ above -- see the SI for further details. 

\emph{Discussion.-}
%
In this work we suggested a phenomenological model which is able to describe partial size control within a broad class of control strategies, and interpolate between the case of constant time to division and division at a critical size, for both size-additive and time-additive noise. We are able to analytically calculate the size and inter-division time distributions for the case of symmetric division, relevant to various bacteria. For \emph{E. coli}, we have shown that a simple biophysical model in which a constant volume is added from consequent events of initiation of DNA replication predicts: 1) Cell size depends exponentially on growth rate. 2) Cell size distributions are approximately log-normal. 3) The coefficient of correlation between size at birth and division is approximately 1/2. 4) The ratio of the CV of size and inter-division time distributions is approximately $\log(2)$.  The simplicity of a biophysical model which implements this idea \cite{incrementalmodel} suggests that this may be a robust way of regulating cell size and coupling DNA replication and growth.

This interpretation in terms of the incremental model suggests an outstanding puzzle: can we underpin the precise molecular mechanism responsible for volume integration? Can the source of the noise in inter-division times be elucidated?  Testing this model further in other microorganisms may yield important insights into cell size regulation, and in particular, it is intriguing to see if the same ideas are applicable to cell size control in higher organisms. Recently, size distributions in other microorganisms were shown to obey simple scaling laws \cite{scaling_lognormal}, suggesting this to be a promising direction, and that the model discussed here may have a broader range of applicability.

\emph{Acknowledgments.-}  This research was supported by the Harvard Society of Fellows and the Milton Fund. We thank Sriram Chandrasekaran, Suckjoon Jun, Andrew W. Murray, Johan Paulsson, Ilya Soifer and Sattar Taheri-Araghi for useful discussions, and to Johan Paulsson for introducing the incremental model to us. We also thank the referees for extremely useful suggestions.
%

%
\clearpage

\section {Cell size regulation in microorganisms - Supplementary Information}

\vspace{0.25 cm}
\textbf{Comparison of theory and numerics for time-additive noise}
\vspace{0.25 cm}

In this section we elaborate on the generality of the results derived in the main text by showing the excellent agreement between the analytical form of Eqs. (5) and (6) of the main text and the numerics. Throughout the SI, the distributions were evaluated numerically by following the lineage of $10^7$ divisions for each case.

%

\begin{figure}[h!]
\includegraphics[width=0.8\linewidth]{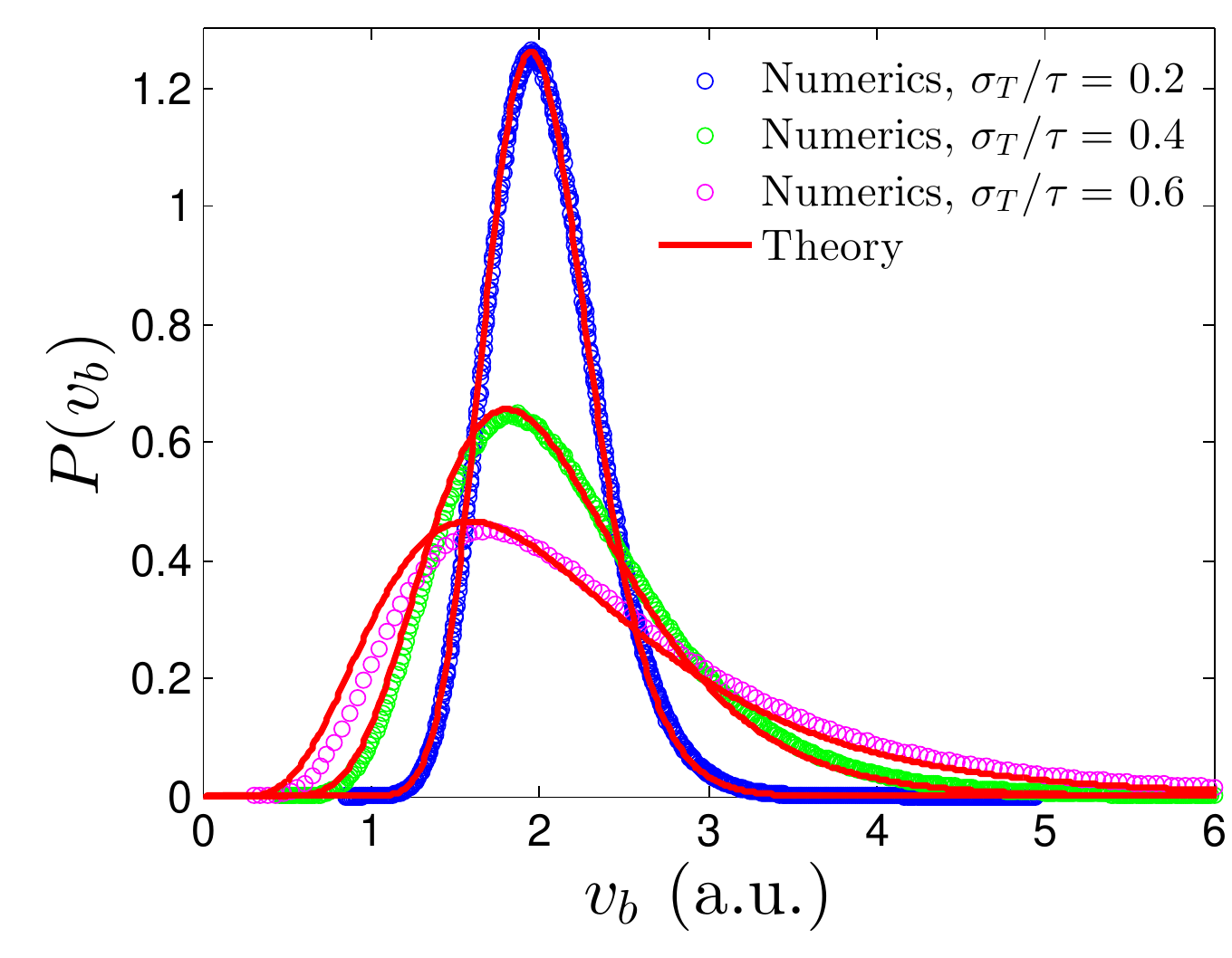}
\caption{Numerical simulation compared with Eq. (5) of the main text, for the incremental model (corresponding to $\alpha=1/2$), for varying noise. The noise is added to the inter-division time and has a standard deviation $\sigma_T$. For each case, the numerical distribution is extracted from a sequence of $10^7$ divisions.}
\label{varying_noise}
\end{figure}

 Fig. \ref{varying_noise} compares the analytical results for $\alpha=1/2$ with numerical results on the incremental model, for varying noise. We tested noise magnitudes ranging from the biologically relevant value of $\sigma_T/\tau=0.2$ to relatively large noise with $\sigma_T/\tau=0.6$. We find excellent agreement even for large noise -- despite the fact that the incremental model is equivalent to the $\alpha=1/2$ analytically tractable model only to lowest order in $(v_b-v_0)$.

\begin{figure}[h!]
\includegraphics[width=0.8\linewidth]{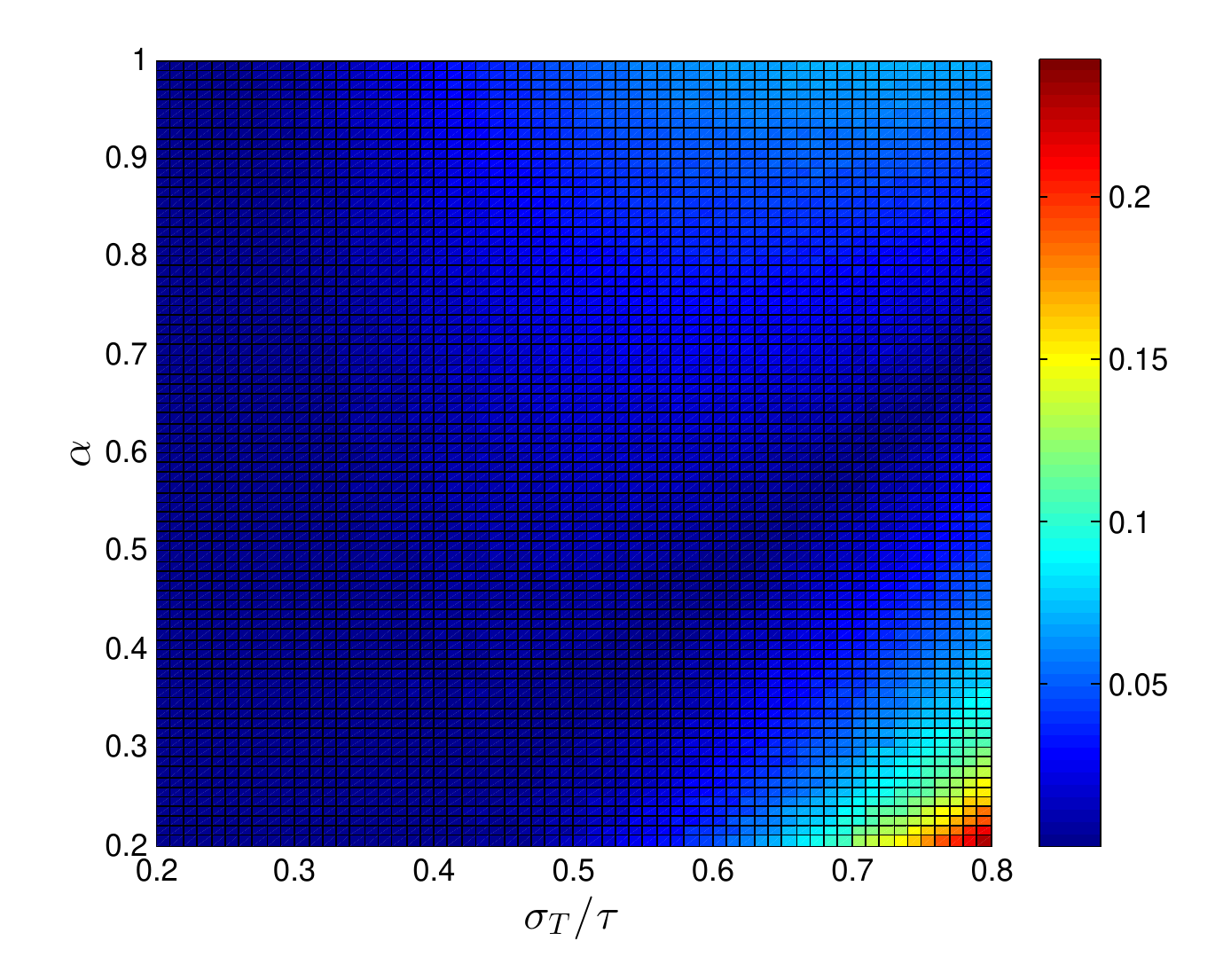}
\caption{The relative error when comparing the theoretical standard deviations (see Eqs. (5) and (6) of the main text), with that found by numerical simulations, where $10^7$ divisions are used to determine each square of the 2d map. Except for non-biologically large magnitudes of noise, the theory captures the numerical results extremely well.}
\label{varying_both}
\end{figure}

Fig. \ref{varying_both} shows the relative error of the standard deviation given by the analytical formula (Eq. (6) of the main text) and the numerics, as both $\alpha$ and the noise are varied. The analytical solution allows for negative growth times (since for large noise the Gaussian time distribution will have a significant negative tail), while in the numerics if the growth time $t_a+t_n$ becomes negative due to anomalously large noise, we replace it with zero -- which is the source of the small relative error between the theory and the numerics. As discussed in the main text, the relative noise is expected to be of the order of 20-30 percent of the doubling time \cite{Xsize1}. For noise of that magnitude, the theory agrees very well with the numerics, for all values of $\alpha$. In fact, it is found that a significant error only occurs for much larger noise and small (unrealistic) values of $\alpha$, leading to very broad size distributions that are never observed biologically.

\vspace{0.25 cm}
\textbf{Solution of model for size-additive noise}
\vspace{0.25 cm}

Using the tools presented in the main text, it is straightforward to modify the approach for size-additive noise.
In this case, it is convenient to use:
\be f(v)=2v_0+2(1-\alpha)(v-v_0). \label{add} \ee

This agrees to lowest order in $(v-v_0)$ with the definition of $\alpha$ of the main text, and hence for the biologically relevant narrow distributions will be essentially equivalent. $\alpha=0$ has no size control, $\alpha=1$ realizes division at a constant size, and most importantly, $\alpha=1/2$ is an \emph{exact} realization of the incremental model.

Now we assume that the noise is added to size rather than time, hence:

\be v^{new}_b =\frac{1}{2} f(v_b)+\xi ,\ee
with $\xi$ a Gaussian variable with standard deviation $\sigma_S$. Using Eq. (\ref{add}) we have:

\be v^{new}_b =\alpha v_0 +(1-\alpha)v_b +\xi, \ee

which can be rewritten as:

\be (v^{new}_b-v_0) = (1-\alpha)(v_b-v_0) +\xi, \ee

Following the same reasoning used in the main text, $v_b-v_0$ will be a Gaussian variable with vanishing mean, and:

\be Var[v_b]=(1-\alpha)^2 Var[v_b]+\sigma_S^2, \ee

Hence:

\be Var[v_b]=\frac{\sigma_S^2}{\alpha(2-\alpha)}. \label{add} \ee

\begin{figure}[t]
\includegraphics[width=0.8\linewidth]{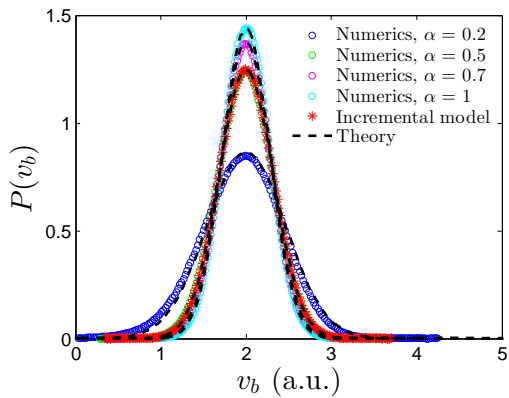}
\caption{Comparison between the analytical results of a Gaussian distribution with variance given by Eq. (\ref{add}) and numerics with \emph{size-additive} noise, for varying values of $\alpha$. Choosing $\alpha=1/2$ provides an exact realization of the incremental model.}
\label{additive_noise}
\end{figure}

Fig. \ref{additive_noise} verifies this result, comparing the analytic results for various values of $\alpha$ with numerics (including $\alpha=1/2$ which corresponds to the incremental model), all with size-additive noise with $\sigma_S/v_0=0.27$ (chosen to capture the width of the experimentally observed distribution).

\vspace{0.25 cm}
\textbf{Distinguishing between size-additive and time-additive noise}
\vspace{0.25 cm}

\begin{figure}[t]
\includegraphics[width=0.8\linewidth]{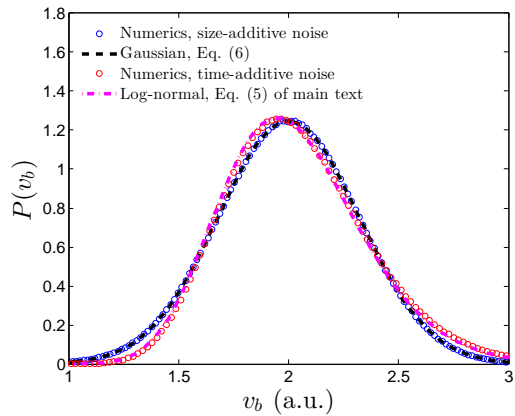}
\caption{The distribution of newborn cell volume was found numerically, using the incremental model for both the cases of size-additive and time-additive noise. In the first case $\sigma_S/v_0=0.2 \log(2)\approx 0.14 $, while in the latter $\sigma_T/\tau_d=0.2$. As is shown, the two cases are well-approximated by Gaussian and log-normal distributions, respectively, and by the theory corresponding to Eq. (\ref{add}) of the SI and of Eq. (5) of the main text.}
\label{dist_size_4}
\end{figure}

For the realistic biological parameters, the difference between the time-additive and size-additive noise is not dramatic, as is shown in Fig. \ref{dist_size_4}.

%

We also tested the size distributions resulting from the regulation strategy corresponding to a generic value of $\alpha$ (Eq. (3) in the main text), with a noise that has \emph{both} a time-additive component (with standard deviation $\sigma_T$) and a size-additive component (with standard deviation $\sigma_S$). Based on the results described above that an approximately Gaussian distribution occurs for a size-additive noise, we expect that only the time-additive component of the noise will significantly contribute to the skewness of the distribution, since a Gaussian distribution has vanishing skewness. Therefore, we calculated numerically the skewness of the newborn size distribution: $\gamma_1 = \mathbb{E}(v_b - \mu)^3/\sigma^3$, where  $\mu \approx v_0$ is the average newborn size and $\sigma$ is the standard deviation of the size distribution.

\begin{figure}[h!]
\includegraphics[width=0.8\linewidth]{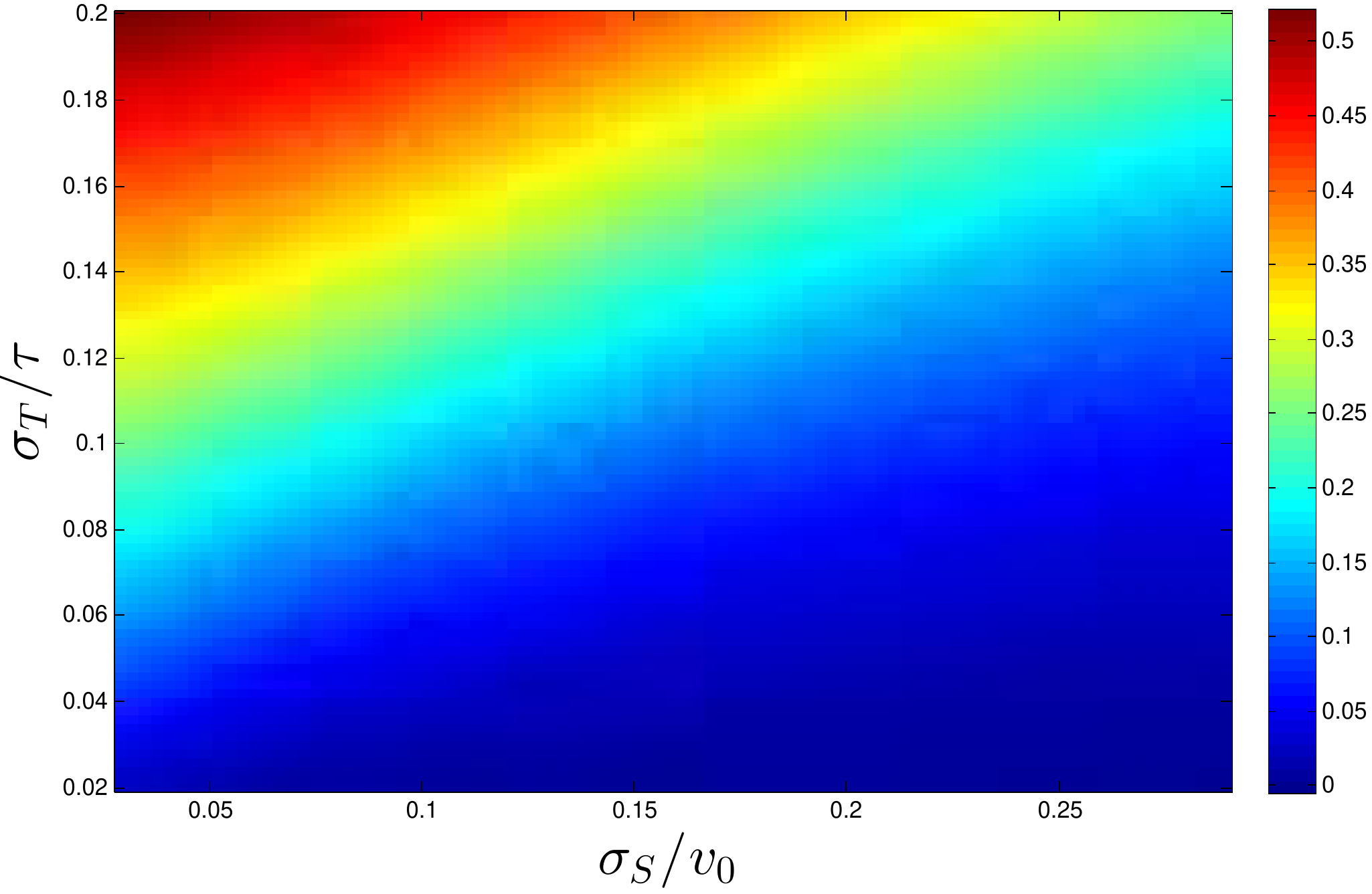}
\caption{The skewness of the newborn size distribution is shown in the 2d map for the mixed case where both time-additive and size-additive noise are present, with magnitudes $\sigma_T$ and $\sigma_S$. As is shown, distributions with non-negligible skewness only occur in the presence of time-additive noise.}
\label{skew}
\end{figure}

As expected, we found that in order to have significant skewness, there has to be a time-additive noise component present, as is illustrated in Fig. \ref{skew}.

Interestingly, in the analysis of financial markets a similar strategy is used to test the multiplicative nature of the stochastic processes underlying the market fluctuations \cite{Xeconophysics_book, Xbouchaud_potters}.

\vspace{0.25 cm}
\textbf{Experimental support for time-additive noise}
\vspace{0.25 cm}

The purpose of this section is to show that existing experimental data for \emph{E. coli} supports time-additive (i.e., multiplicative) noise rather than size-additive noise, using the tool developed in the previous section: namely, the skewness of the size distribution.

Ref. [\onlinecite{Xcorrelation}] measures the newborn size distribution of \emph{E. coli} using the membrane elution technique, and shows that it has non-negligible positive skewness and it agrees very well with a log-normal distribution.
%

Recently, Ref. [\onlinecite{Xsize1}] analyzed cell size fluctuations and distributions. Their size distributions are strongly skewed to the right ($\gamma_1 > 0.5$), suggesting that the noise is multiplicative rather than additive to the size. A similar analysis was performed in Ref. [\onlinecite{Xsize2}], which further corroborates our conclusion.

\begin{figure}[t]
\includegraphics[width=1\linewidth]{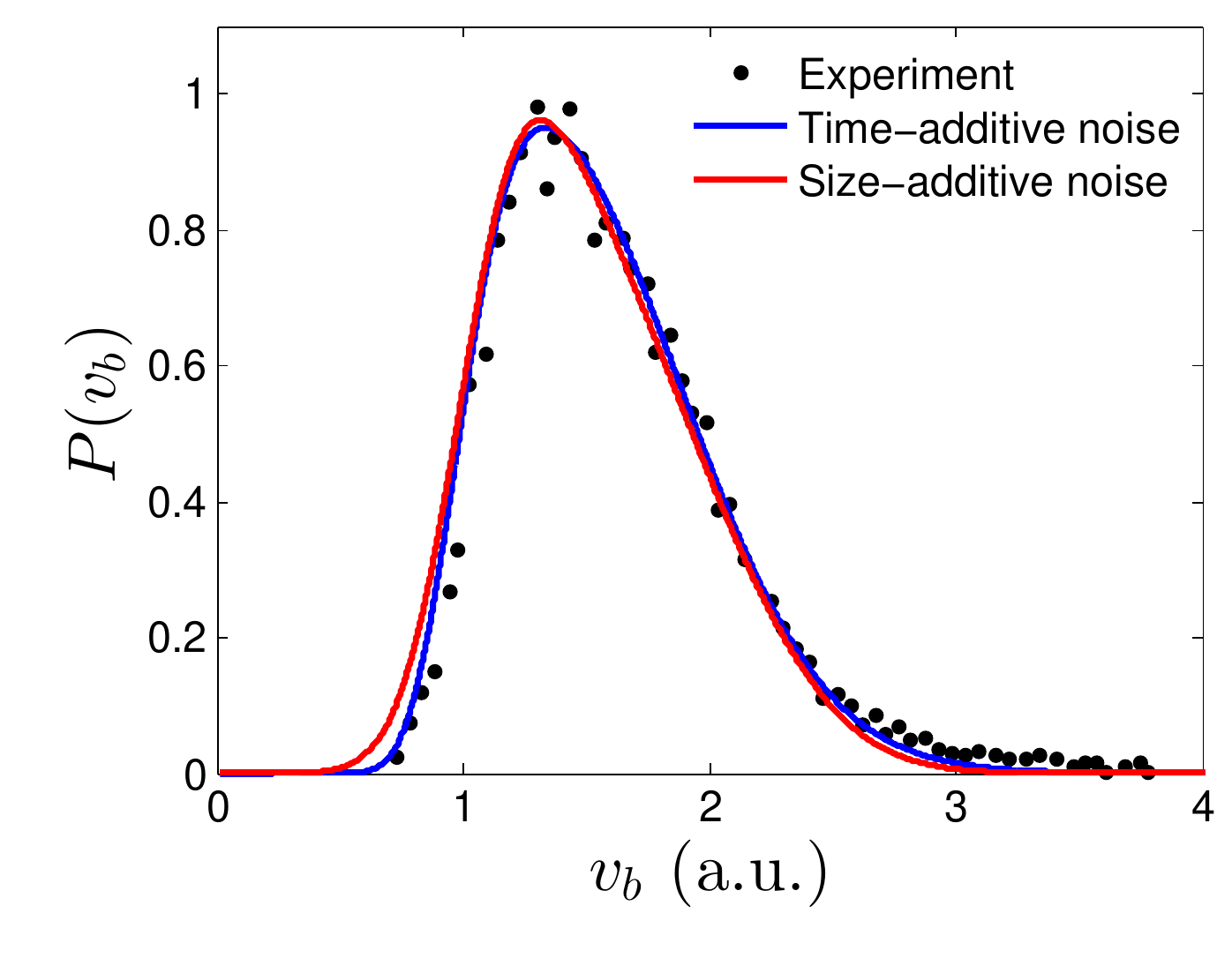}
\caption{The distribution of cell volume measured in \emph{E. coli} cells (from Ref. [\onlinecite{Xkubitschek}].) agrees well with a log-normal newborn size distribution. The skewness of the distribution is consistent with a time-additive noise, and is much larger than that obtained with a size-additive noise.}
\label{dist_size}
\end{figure}

In various other experimental setups the distribution over an \emph{entire} population of cells at various stages of the cell cycle is measured, for example by taking single microscopy images (note that this is only relevant for size distributions, and not for the inter-division time distribution). We have found that the resulting size distribution over the entire population (which is also well approximated by a log-normal distribution), agrees well with existing experimental data, see Fig. \ref{dist_size}, obtained for \emph{E. coli} in slow growth conditions. We also compare the data with the size distribution resulting from a similar model to the one we discussed thus far, but with a size-additive noise rather than a time-additive noise. Note that since the distribution is in this case approximately a sum of shifted Gaussians, it will not have the vanishing skewness as the newborn size distribution does for size-additive noise. Nevertheless, we find that in this case the skewness of the measured distribution is compatible with a time-additive noise but much larger than that expected for size-additive noise (0.3 compared with approximately 1 for the experimentally observed distribution). 

Therefore, these experiments support a time-additive rather than size-additive noise.

Although these various experiments suggest that noise in the size-control of bacteria is multiplicative rather than additive to size, a significant amount of additional experimental work is needed in order to elucidate the origin of the noise at the molecular level. Ref. [\onlinecite{Xsiggia}] used the powerful tools of molecular biology to gain insights into precisely such a problem in budding yeast. In this pioneering work, they also model the noise as time-additive rather than size-additive, and are able to deduce the molecular mechanism for it and \emph{modify it} in order to test their stochastic model. It would be highly rewarding to do the same for bacteria.

\vspace{0.25 cm}
\textbf{DnaA as a volume integrator}
\vspace{0.25 cm}

Here, we review some recent progress in our understanding of the function of dnaA in initiating DNA replication in bacteria, and show that it shares many properties with the hypothetical model described in the main text.
DnaA has two forms, an active, ATP-bound form, and an inactive ADP-bound form.
It is believed that 20 dnaA proteins in their active form are needed in order to initiate DNA replication \cite{Xlevin, Xdonachie2}. This highly cooperative mechanism is parallel to the ``critical number" of protein $B$ in the main text ($P_2$ in Ref. \cite{Xincrementalmodel}). After initiation, the number of active copies of dnaA quickly drops \cite{Xdonachie2}, which is equivalent to the ``degradation" of protein $B$.

Importantly, dnaA is known to autoregulate \cite{Xwright}, which in certain cases leads to a concentration approximately independent of the growth rate \cite{Xwright2} -- this was the requirement for protein $A$ of the main text ($P_1$ in Ref. \cite{Xincrementalmodel}).

As expected, inducing high levels of dnaA in \emph{E. coli} leads to early initiation \cite{Xdnaa}, while dnaA mutants initiate at a later time \cite{Xkleckner,Xkleckner2}.

Therefore, it is possible that the two forms of dnaA serve together to implement the volume integration which is hypothetically discussed in Ref. \cite{Xincrementalmodel}. Certainly, many other proteins are involved in the process, such as SeqA \cite{Xkleckner}, and further experiments are needed to better understand the biochemical and biophysical mechanisms. Nevertheless, the results presented in the main text lead to severe constraints on possible models.

\end{document}